\documentclass[12pt]{article}
\usepackage{amsmath,amssymb,latexsym}

\begin{document}

\begin{titlepage}
\noindent

\begin{flushright}
	YITP-02-33 \\
	hep-th/0205054
\end{flushright}

\vspace{2.5cm}

\begin{center}
{\Large\bf Supersymmetric Field Theories}\\
\vspace{0.5cm}
{\Large\bf on Noncommutative Spaces}\\
\bigskip
\vspace{1.5cm}
{\large Yoshinobu Habara}~\footnote{e-mail: habara@yukawa.kyoto-u.ac.jp}\\
\vspace{0.5cm}
{\it  Yukawa Institute for Theoretical Physics,}\\
{\it  Kyoto University, Kyoto 606-8502, Japan}
\end{center}

\vspace{1.5cm}

\begin{abstract}

Supersymmetric field theories on noncommutative spaces are constructed. We present two different representations of noncommutative space, but we can obtain supersymmetry algebla and supersymmetric Yang-Mills action independent of its representation. As a result, we will see that the action has a close relationship with IIB matrix model.

\end{abstract}

\vspace{0.5cm}

\end{titlepage}

%%%%%%%%%%%%%%%%%%%%%%%%%%%%%%%%%%%%%%%%%%%%%%%%%%%%%%%%%%%%%%%%%%%%%%%%%%%%%%%
\noindent
{\bf 1. Introduction}

\vskip 0.5cm

The extended geometry known as noncommutative geometry has attracted many physists' attentions for years (see, for example,~\cite{schwarz},~\cite{susskind},~\cite{nekrasov}, and~\cite{witten} etc.), because the unusual commutation relation about space-time coordinates $[\hat{x}^{\mu},\hat{x}^{\nu}]=i\Theta^{\mu \nu}$ may regularize the divergence arises from short distance of space-time, especially in gravity model, as the commutation relation $[\hat{x}^{\mu},\hat{p}_{\nu}]=i\hbar \delta^{\mu}_{\> \> \nu}$ leads $\Delta x\Delta p\geq \hbar$. The notion of noncommutative geometry is that we regard the ``point" $x^{\mu}$ in noncommutative space as the operator $\hat{x}^{\mu}$ acting on a certain Hilbert space $\mathcal{H}$, and the eigenvalues of $\hat{x}^{\mu}$ as the space-time coordinates (see~\cite{connes}and~\cite{landi}). In order to include the space-time noncommutativity into various field theories, the most well-known method is the $*$-product formalism, but we would not employ it here.

In this paper, we would like to adopt more algebraic method because of its beautifulness. Our approach is based on the argument that first quantization is equivalent to the introduction of noncommutativity into the phase space. We denote such a noncommutative space $\mathcal{A}$. Following this argument, we can incorporate gauge field in a usual way, and we find a Yang-Mills action on $\mathcal{A}$ is very like the bosonic part of IIB matrix model, so we also want to investigate its fermionic part. Therefore, we try to construct supersymmetry algebra on noncommutative space $\mathcal{A}$. We see that the algebra goes successfully in constructing supersymmetric scalar field theory, so is not very wrong. With this supersymmetry algebra, the Yang-Mills action can be obtained in the same way as the ordinary supersymmetry algebra through vector superfield, independent of the representation of noncommutative space $\mathcal{A}$ (surprisingly !!). The supersymmetric Yang-Mills action on noncommutative space $\mathcal{A}$ resembles IIB matrix model, and is equivalent to in $10$-dimension, i.e. when noncommutaive-space algebra $\mathcal{A}\oplus End_{\mathcal{A}}\mathcal{H}$ is generated from $10+10$ operators. And, with their relationship, we can evaluate the parameter $\Theta$ which expresses space-time noncommutativity, and will find that $\Theta$ must be at the same order as the string length $\alpha^{\prime}$.

The organization of this paper is as follows. \S 2 is devoted to the review of~\cite{habara}, the representation of noncommutative space $\mathcal{A}$ and scalar field theory on it, and the introduction of the gauge field. In \S 3, we construct the supersymmetry algebra on noncommutative space $\mathcal{A}$ and supersymmetric scalar field theory on it by using the representation presented in \S 2 which reproduces scalar field theory in \S 2. And in \S 4, with the results in \S 3, supersymmetric Yang-Mills theory is obtained and we evaluate the noncommutative parameter $\Theta$ in comparison with string theory in special gauge, i.e. IIB matrix model. In \S 5, discussions are presented.

\vskip 1.5cm

%%%%%%%%%%%%%%%%%%%%%%%%%%%%%%%%%%%%%%%%%%%%%%%%%%%%%%%%%%%%%%%%%%%%%%%%%%%%%%%
\noindent
{\bf 2. Scalar field theory and introduction of gauge field}

\vskip 0.5cm

The purpose of this section is to introduce gauge fields into field theory on noncommutative space. To do this, let us start with the review of~\cite{habara}, the representation of noncommutative space and the corresponding scalar field theory.

First, we present an algebra $\mathcal{A}$ of noncommutative space: 

\begin{eqnarray}
	& & [\hat{x}^{\mu}, \hat{x}^{\nu}]=i\Theta^{\mu \nu}, \nonumber \\
	& & [\hat{x}^{\mu}, \hat{p}_{\nu}]=i\hbar \delta^{\mu}_{\> \> \nu}, 
	\qquad \quad \mu ,\nu =0,1,\cdots ,(d-1) \\
	& & [\hat{p}_{\mu}, \hat{p}_{\nu}]=-i\hbar^2 \Theta^{-1}_{\mu \nu}. 
	\nonumber \\
	& & \nonumber \\
	& & \qquad \eta^{\mu \nu}=\text{diag}(-1,+1,\cdots ,+1) \nonumber \\
	& & \qquad \hbar^2 \ll \Theta \ll \hbar \nonumber
\end{eqnarray}

\noindent In the following, we set $\hbar =1$. This algebra has a representation as operators acting on the Hilbert space $\mathcal{H}$ consisting of square-integrable functions. That is 

\begin{eqnarray}
	& & \left\{ 
	\begin{array}{l}
	\hat{x}^{\mu}=\displaystyle \frac{1}{2}x^{\mu}+i\Theta^{\mu \nu}
	\partial_{\nu} \\
	\hat{p}_{\mu}=-i\partial_{\mu}-\displaystyle \frac{1}{2}
	\Theta^{-1}_{\mu \nu} x^{\nu} \end{array} \right. ,
	\qquad \hat{p}_{\mu}=-\Theta^{-1}_{\mu \nu}\hat{x}^{\nu} \\
	& & \qquad \text{acting on } f(x)=f(x^0,x^1,\cdots ,x^{d-1}) \in 
	\mathcal{H}. \nonumber
\end{eqnarray}

\noindent This representation space is $d$-dimensional, and it seems to be too many. In fact, for example, if $d=4$, we can represent the algebra on $2$-dimensional space: 

\begin{eqnarray*}
	& & \hat{x}^0=x^0+\partial_1, \quad \hat{x}^1=x^0+x^1, \\
	& & \hat{x}^2=x^1+\partial_0, \quad \hat{x}^3=\partial_0+\partial_1.
\end{eqnarray*}

\noindent But in our representation (2), as we shall see later, we must include the algebra $End_{\mathcal{A}}\mathcal{H}$, so the $d$-dimensional space is needed.

Next, we will construct scalar field theory on the noncommutative space $\mathcal{A}$. This model is constructed from the Einstein energy-momentum relation $p_{\mu}p^{\mu}+m^2=0$ by replacing the phase-space coordinates with the corresponding operators: $x^{\mu}, p_{\mu}\! \to \! \hat{x}^{\mu}, \hat{p}_{\mu}$.

\begin{eqnarray}
	& & (\hat{p}_{\mu}\hat{p}^{\mu}+m^2)\phi(x)=0 \\
	& & \Big \{ (-i\partial_{\mu}-\frac{1}{2}\Theta^{-1}_{\mu \nu}x^{\nu})
	(-i\partial^{\mu}-\frac{1}{2}\Theta^{-1 \mu \rho}x_{\rho})+m^2\Big \} 
	\phi(x)=0 \nonumber
\end{eqnarray}

\noindent The action that produces this equation of motion (3) is 

\begin{eqnarray}
	S \! \! \! \! \! \! \! 
	& & =\int \! d^d x \phi^{\dagger}(x) \Big \{ (-i\partial_{\mu}
	-\frac{1}{2}\Theta^{-1}_{\mu \nu}x^{\nu})(-i\partial^{\mu}-\frac{1}{2}
	\Theta^{-1 \mu \rho}x_{\rho})+m^2\Big \} \phi (x) \nonumber \\
	& & =\int \! d^d x \Big \{ -(-i\partial_{\mu}+\frac{1}{2}
	\Theta^{-1}_{\mu \nu}x^{\nu})\phi^{\dagger}(x) (-i\partial^{\mu}
	-\frac{1}{2}\Theta^{-1 \mu \rho}x_{\rho})\phi (x)+m^2 \phi^{\dagger}(x)
	\phi (x)\Big \}.
\end{eqnarray}

\noindent From this action, we obtain the equation of motion for $\phi^{\dagger}(x)$ at the same time: 

\begin{eqnarray}
	& & \Big \{ (-i\partial_{\mu}+\frac{1}{2}\Theta^{-1}_{\mu \nu}x^{\nu})
	(-i\partial^{\mu}+\frac{1}{2}\Theta^{-1 \mu \rho}x_{\rho})+m^2\Big \} 
	\phi^{\dagger}(x)=0. \nonumber \\
	& & \qquad \Uparrow \nonumber \\
	& & (\hat{p}^{\prime}_{\mu}\hat{p}^{\prime \mu}+m^2)\phi^{\dagger}(x)=0
\end{eqnarray}

\noindent The operators $\hat{x}^{\prime \mu}$ and $\hat{p}^{\prime}_{\mu}$ generate $End_{\mathcal{A}}\mathcal{H}$, i.e. $\mathcal{A}$-linear maps $\rho :\mathcal{H}\! \to \! \mathcal{H}, \> \rho (af)=a\rho (f)$ for $a\in \mathcal{A}, f\in \mathcal{H}$.

\begin{eqnarray}
	& & [\hat{x}^{\prime \mu}, \hat{x}^{\prime \nu}]=-i\Theta^{\mu \nu}, 
	\nonumber \\
	& & [\hat{x}^{\prime \mu}, \hat{p}^{\prime}_{\nu}]
	=i\delta^{\mu}_{\> \> \nu}, \\
	& & [\hat{p}^{\prime}_{\mu}, \hat{p}^{\prime}_{\nu}]
	=i\Theta^{-1}_{\mu \nu}. \nonumber \\
	& & \nonumber \\
	& & \left\{ 
	\begin{array}{l}
	\hat{x}^{\prime \mu}=\displaystyle \frac{1}{2}x^{\mu}-i\Theta^{\mu \nu}
	\partial_{\nu} \\
	\hat{p}^{\prime}_{\mu}=-i\partial_{\mu}+\displaystyle \frac{1}{2}
	\Theta^{-1}_{\mu \nu} x^{\nu} \end{array} \right. ,
	\qquad \hat{p}^{\prime}_{\mu}=\Theta^{-1}_{\mu \nu}
	\hat{x}^{\prime \nu}.
\end{eqnarray}

\noindent All $\hat{x}^{\prime \mu}$ and $\hat{p}^{\prime}_{\mu}$ are commuting with $\hat{x}^{\mu}$ and $\hat{p}_{\mu}$.

We are now ready to introduce gauge fields. Here, we also use the same method as the previous scalar field theory. That is, we will replace the phase-space coordinates $x^{\mu},p_{\mu}$ and gauge field $A_{\mu}(x)$ in the Einstein energy-momentum relation in the presence of gauge field with the corresponding operators acting on the Hilbert space $\mathcal{H}$.

To begin with, let us define the derivation $\delta_{\mu}$ in $\mathcal{A}$: 

\begin{eqnarray*}
	& & \delta_{\mu}a\equiv [i\hat{p}_{\mu}, a], \qquad a\in \mathcal{A} \\
	& & \quad \Rightarrow [\delta_{\mu},\delta_{\nu}]=0, \quad 
	\delta_{\mu} \hat{x}^{\nu}=\delta_{\mu}^{\> \> \nu}.
\end{eqnarray*}

\noindent By making use of this derivation, we introduce the gauge field $\hat{A}_{\mu}$ and the connection $\nabla^{\prime}_{\mu}$ which satisfies Leibnitz rule for $a\in \mathcal{A}, f\in \mathcal{H}$: 

\begin{eqnarray}
	& & (p_{\mu}+A_{\mu})(p^{\mu}+A^{\mu})+m^2=0 \nonumber \\
	& & \qquad \Downarrow \nonumber \\
	& & \Big \{ (\hat{p}_{\mu}+\hat{A}^{\prime}_{\mu})(\hat{p}^{\mu}
	+\hat{A}^{\prime \mu})+m^2 \Big \}\phi (x)=0, \\
	& & \nonumber \\
	\text{Leibnitz} \! \! \! \! \! \! \! 
	& & \text{rule} \nonumber \\
	& & \nabla^{\prime}_{\mu}(af)\equiv (i\hat{p}_{\mu}
	+i\hat{A}^{\prime}_{\mu})(af) \\
	& & \qquad \quad \; \; =(\delta_{\mu}a)f+a\nabla^{\prime}_{\mu}f. 
	\nonumber
\end{eqnarray}

\noindent From this equation, we obtain 

\begin{eqnarray*}
	& & [\hat{A}^{\prime}_{\mu},a]=0, \quad \forall a\in \mathcal{A} \\
	& & \hat{A}^{\prime}_{\mu}\in End_{\mathcal{A}}\mathcal{H}.
\end{eqnarray*}

\noindent Therefore, we write $\hat{A}^{\prime}_{\mu}=A_{\mu}(\hat{x}^{\prime})$. Similarly, 

\begin{eqnarray}
	& & \Big \{ (\hat{p}^{\prime}_{\mu}-\hat{A}_{\mu})(\hat{p}^{\prime \mu}
	-\hat{A}^{\mu})+m^2 \Big \}\phi^{\dagger}(x)=0, \\
	& & \nabla_{\mu}\equiv i\hat{p}^{\prime}_{\mu}-i\hat{A}_{\mu}, \\
	& & \hat{A}_{\mu}=A_{\mu}(\hat{x})\in \mathcal{A}. \nonumber
\end{eqnarray}

\noindent As we have seen, $\hat{p}_{\mu}\in \mathcal{A}$ couples to $A_{\mu}(\hat{x}^{\prime})\in End_{\mathcal{A}}\mathcal{H}$, and $\hat{p}_{\mu}^{\prime}\in End_{\mathcal{A}}\mathcal{H}$ to $A_{\mu}(\hat{x})\in \mathcal{A}$, so we must consider $\mathcal{A}\oplus End_{\mathcal{A}}\mathcal{H}$ as the algebra of noncommutative space.

The action that produces the equations of motion (8)(10) is 

\begin{eqnarray}
	S=\int \! d^d x \Big \{ -(\hat{p}^{\prime}_{\mu}-A_{\mu}(\hat{x}))
	\phi^{\dagger}(x) (\hat{p}^{\mu}+A^{\mu}(\hat{x}^{\prime}))\phi (x)
	+m^2 \phi^{\dagger}(x)\phi (x)\Big \}.
\end{eqnarray}

\noindent The gauge transformation is realized in the algebra $\mathcal{A}$ and $End_{\mathcal{A}}\mathcal{H}$ respectively. For any unitary element $U(\hat{x})\in \mathcal{A}$ and $U(\hat{x}^{\prime})\in End_{\mathcal{A}}\mathcal{H}$,

\begin{eqnarray}
	& & \nabla_{\mu} \to U^{\dagger}(\hat{x}) \nabla_{\mu} U(\hat{x}) 
	\quad \Rightarrow \quad A_{\mu}(\hat{x}) \to U^{\dagger}(\hat{x}) 
	A_{\mu}(\hat{x}) U(\hat{x}), \\
	& & \nabla^{\prime}_{\mu} \to U^{\dagger}(\hat{x}^{\prime}) \nabla_{\mu}
	 U(\hat{x}^{\prime})\quad \Rightarrow \quad A_{\mu}(\hat{x}^{\prime}) 
	\to U^{\dagger}(\hat{x}^{\prime}) A_{\mu}(\hat{x}^{\prime}) 
	U(\hat{x}^{\prime}).
\end{eqnarray}

\noindent And the gauge invariance of tha action (12) is achieved by the partial integration under the transformations $\phi (x) \! \to \! U(\hat{x}^{\prime})\phi (x)$ and $\phi^{\dagger}(x) \! \to \! U^{\dagger}(\hat{x})\phi^{\dagger}(x)$.

From the connections (9)(11), we can construct the field strength $\hat{F}_{\mu \nu}\in \mathcal{A}$ and $\hat{F}^{\prime}_{\mu \nu} \in End_{\mathcal{A}}\mathcal{H}$, respectively, in the following way.

\begin{eqnarray}
	\hat{F}_{\mu \nu} \! \! \! \! \! \! 
	& & \equiv [\nabla_{\mu}, \nabla_{\nu}] \nonumber \\
	& & =-[\hat{p}^{\prime}_{\mu}-A_{\mu}(\hat{x}), \hat{p}^{\prime}_{\nu}
	-A_{\nu}(\hat{x})] \nonumber \\
	& & =-[A_{\mu}(\hat{x}), A_{\nu}(\hat{x})]-i\Theta^{-1}_{\mu \nu}, \\
	\hat{F}^{\prime}_{\mu \nu} \! \! \! \! \! \! 
	& & \equiv [\nabla^{\prime}_{\mu}, \nabla^{\prime}_{\nu}] \nonumber \\
	& & =-[\hat{p}_{\mu}+A_{\mu}(\hat{x}^{\prime}), \hat{p}_{\nu}
	+A_{\nu}(\hat{x}^{\prime})] \nonumber \\
	& & =-[A_{\mu}(\hat{x}^{\prime}), A_{\nu}(\hat{x}^{\prime})]
	+i\Theta^{-1}_{\mu \nu}.
\end{eqnarray}

\noindent And their gauge transformations are 

\begin{eqnarray}
	& & \hat{F}_{\mu \nu} \to U^{\dagger}(\hat{x}) \hat{F}_{\mu \nu} 
	U(\hat{x}), \\
	& & \hat{F}^{\prime}_{\mu \nu} \to U^{\dagger}(\hat{x}^{\prime}) 
	\hat{F}^{\prime}_{\mu \nu} U(\hat{x}^{\prime}).
\end{eqnarray}

\noindent After we construct the integral, in other words, the trace which has cyclic symmetry $\displaystyle \int \! ab=\int \! ba$ (see Appendix), we can obtain the Yang-Mills action on noncommutative space $\mathcal{A}\oplus End_{\mathcal{A}}\mathcal{H}$: 

\begin{eqnarray}
	S_{YM} \! \! \! \! \! \! 
	& & =\int \hat{F}_{\mu \nu} \hat{F}^{\mu \nu}+\int 
	\hat{F}^{\prime}_{\mu \nu} \hat{F}^{\prime \mu \nu} \nonumber \\
	& & =\int \Big \{ [A_{\mu}(\hat{x}),A_{\nu}(\hat{x})][A^{\mu}(\hat{x})
	,A^{\nu}(\hat{x})] \nonumber \\
	& & \qquad \qquad +[A_{\mu}(\hat{x}^{\prime}),
	A_{\nu}(\hat{x}^{\prime})][A^{\mu}(\hat{x}^{\prime}),
	A^{\nu}(\hat{x}^{\prime})]-2\Theta^{-1}_{\mu \nu} \Theta^{-1 \mu \nu} 
	\Big \}.
\end{eqnarray}

\noindent Here, we find that the Yang-Mills action on noncommutative space looks very like the bosonic part of IIB matrix model if d=10. Therefore, in the next section, we will try to introduce the supersymmetry algebra on noncommutative space, and investigate the fermionic part of the Yang-Mills theory in the later section.

\vskip 1.5cm

%%%%%%%%%%%%%%%%%%%%%%%%%%%%%%%%%%%%%%%%%%%%%%%%%%%%%%%%%%%%%%%%%%%%%%%%%%%%%%%
\noindent
{\bf 3. Noncommutative supersymmetry algebra and chiral superfield theory}

\vskip 0.5cm

In this section, let us give a supersymmetry algebra on noncommutative space, and supersymmteric chiral field theory. We will see that, under the representation (2)(7), the bosonic part of the supersymmetric action of chiral superfield reproduces the action (4).

For concrete calculation, we set $d\! =\! 4$, i.e. $\mu ,\nu ,\rho ,\lambda ,\cdots$ are $SO(3,1)$ vector indices and $\alpha ,\beta ,\gamma ,\delta ,\cdots$ $SO(3,1)$ Weyl spinor indices. $\theta^{\alpha},\bar{\theta}_{\dot{\alpha}}$ are Grassmann coordinates, and the Hilbert space $\mathcal{H}$ is extended to $\mathcal{H}\oplus \{ \theta^{\alpha},\bar{\theta}_{\dot{\alpha}}\}$. For detailed notation, see~\cite{wess}.

First, we give the supersymmetry algebra.

\begin{eqnarray}
	& & \{Q_{\alpha},\bar{Q}_{\dot{\alpha}} \}
	=-\sigma^{\mu}_{\alpha \dot{\alpha}}(\hat{P}_{\mu}
	+\hat{P}^{\prime}_{\mu}) \nonumber \\
	& & \{ Q_{\alpha},Q_{\beta}\} =\{ \bar{Q}_{\dot{\alpha}},
	\bar{Q}_{\dot{\beta}}\}=0 \nonumber \\
	& & [Q_{\alpha},\hat{P}_{\mu}+\hat{P}^{\prime}_{\mu}]
	=[\bar{Q}_{\dot{\alpha}},\hat{P}_{\mu}+\hat{P}^{\prime}_{\mu}]=0 \\
	& & [\hat{P}_{\mu},\hat{P}_{\nu}]=-i\Theta^{-1}_{\mu \nu}, \quad 
	[\hat{P}^{\prime}_{\mu},\hat{P}^{\prime}_{\nu}]=i\Theta^{-1}_{\mu \nu} 
	\nonumber \\
	& & [\hat{P}_{\mu},\hat{P}^{\prime}_{\nu}]=0 \nonumber
\end{eqnarray}

The representation of this algebra, which does not depend on that of $\mathcal{A}\oplus End_{\mathcal{A}}\mathcal{H}$, is 

\begin{eqnarray}
	& & \left\{ \begin{array}{l}
	Q_{\alpha}=\displaystyle \frac{\partial}{\partial \theta^{\alpha}}
	+(\sigma^{\mu} \bar{\theta})_{\alpha} \hat{P}_{\mu} \\
	\bar{Q}_{\dot{\alpha}}
	=-\displaystyle \frac{\partial}{\partial \bar{\theta}^{\dot{\alpha}}}
	-(\theta \sigma^{\mu})_{\dot{\alpha}} \hat{P}^{\prime}_{\mu} 
	\end{array} \right. ,\\
	& & \nonumber \\
	& & \left\{ \begin{array}{l}
	\hat{P}_{\mu}=\hat{p}_{\mu}-\displaystyle \frac{i}{2}
	\Theta^{-1}_{\mu \nu}(\theta \sigma^{\mu}\bar{\theta}) \\
	\hat{P}^{\prime}_{\mu}=\hat{p}^{\prime}_{\mu}
	-\displaystyle \frac{\mathstrut i}{2}
	\Theta^{-1}_{\mu \nu}(\theta \sigma^{\mu}\bar{\theta})
	\end{array} \right. .
\end{eqnarray}

\noindent Similarly, let us define the supersymmetric derivatives as follows: 

\begin{eqnarray}
	& & \left\{ \begin{array}{l}
	D_{\alpha}=\displaystyle \frac{\partial}{\partial \theta^{\alpha}}
	-(\sigma^{\mu} \bar{\theta})_{\alpha} \hat{P}^{\prime}_{\mu} \\
	\bar{D}_{\dot{\alpha}}
	=-\displaystyle \frac{\partial}{\partial \bar{\theta}^{\dot{\alpha}}}
	+(\theta \sigma^{\mu})_{\dot{\alpha}} \hat{P}_{\mu}
	\end{array} \right. ,\\
	& & \nonumber \\
	& & \{D_{\alpha},\bar{D}_{\dot{\alpha}} \}
	=\sigma^{\mu}_{\alpha \dot{\alpha}}(\hat{P}_{\mu}
	+\hat{P}^{\prime}_{\mu}), \nonumber \\
	& & \{ D_{\alpha},D_{\beta}\} =\{ \bar{D}_{\dot{\alpha}},
	\bar{D}_{\dot{\beta}}\}=0, \\
	& & [D_{\alpha},\hat{P}_{\mu}+\hat{P}^{\prime}_{\mu}]
	=[\bar{D}_{\dot{\alpha}},\hat{P}_{\mu}+\hat{P}^{\prime}_{\mu}]=0, 
	\nonumber \\
	& & \{ Q_{\alpha},D_{\beta}\} =\{ \bar{Q}_{\dot{\alpha}},
	\bar{D}_{\dot{\beta}}\} =\{ Q_{\alpha},\bar{D}_{\dot{\alpha}}\} 
	=\{ \bar{Q}_{\dot{\alpha}},D_{\alpha}\} =0. \nonumber
\end{eqnarray}

Next, as the representation of $\mathcal{A}\oplus End_{\mathcal{A}}\mathcal{H}$, we adopt (2)(7) to construct the chiral superfield theory, and show that it reproduces the action (4).

Under the representation (2)(7), the Hilbert space is the space of square-integrable functions $F(x,\theta ,\bar{\theta})$ on the superspace $(x^{\mu},\theta^{\alpha} ,\bar{\theta}_{\dot{\alpha}})$. These functions are called superfields and expanded in powers of $\theta ,\bar{\theta}$ as 

\begin{eqnarray}
	& & F(x,\theta ,\bar{\theta})=f(x)+\theta \phi (x)+\bar{\theta}
	\bar{\chi}(x)+\theta \theta m(x)+\bar{\theta}\bar{\theta}n(x) 
	\nonumber \\
	& & \qquad \qquad \qquad \qquad 
	+\theta \sigma^{\mu}\bar{\theta}v_{\mu}(x)
	+\theta \theta \bar{\theta}\bar{\lambda}(x)
	+\bar{\theta}\bar{\theta}\theta \psi (x)
	+\theta \theta \bar{\theta}\bar{\theta}d(x).
\end{eqnarray}

\noindent Let us define the supersymmetry transformations as follows: 

\begin{eqnarray}
	\delta_{\xi}F(x,\theta ,\bar{\theta})\equiv (\xi Q+\bar{\xi}\bar{Q})
	F(x,\theta ,\bar{\theta}).
\end{eqnarray}

\noindent Under these transformations, we can construct the invariant integral. Here, the integrations about the Grassmann coordinates $\theta ,\bar{\theta}$ are regularized.

\begin{eqnarray}
	& & I=\int d^4 x d\theta d\theta d\bar{\theta} d\bar{\theta} \> \> 
	e^{-\frac{1}{2}\theta \sigma^{\mu}\bar{\theta}\Theta^{-1}_{\mu \nu}
	x^{\nu}}F(x,\theta ,\bar{\theta}) \\
	& & \delta_{\xi}I=\int d^4 x\> (\text{total derivatives})=0 
	\nonumber \\
	& & \quad \int d\theta d\theta (\theta \theta )=1,\quad 
	\int d\bar{\theta} d\bar{\theta} (\bar{\theta}\bar{\theta})=1 \nonumber
\end{eqnarray}

Let us turn to the chiral superfields. Similar to the ordinary case, we define the chiral superfields on noncommutative space as 

\begin{eqnarray}
	\bar{D}_{\dot{\alpha}}\Phi (x,\theta ,\bar{\theta})=0.
\end{eqnarray}

\noindent The field $\Phi$ lives in the Hilbert space. The general solution of this equation is 

\begin{eqnarray}
	\Phi(x,\theta ,\bar{\theta}) \! \! \! \! \! \! \! 
	& & =e^{\frac{1}{2}\theta \sigma^{\mu}\bar{\theta}\Theta^{-1}_{\mu \nu}
	x^{\nu}} \{ A(y)+\sqrt{2}\theta \psi (y)+\theta \theta F(y)\},
	\quad  y^{\mu}=x^{\mu}+i\theta \sigma^{\mu}\bar{\theta} 
	\nonumber \\
	& & =A(x)-\theta \sigma^{\mu}\bar{\theta}\hat{p}_{\mu}A(x)
	-\frac{1}{4}\theta \theta \bar{\theta}\bar{\theta}\hat{p}_{\mu}
	\hat{p}^{\mu}A(x) \nonumber \\
	& & \qquad \qquad 
	+\sqrt{2}\theta \psi (x)+\frac{1}{\sqrt{2}}
	\theta \theta \hat{p}_{\mu}\psi (x)\sigma^{\mu}\bar{\theta}
	+\theta \theta F(x).
\end{eqnarray}

\noindent The anti-chiral superfields are defined through the equation: 

\begin{eqnarray}
	D_{\alpha}\Phi^{\dagger}(x,\theta ,\bar{\theta})=0.
\end{eqnarray}

\noindent And its general solution is 

\begin{eqnarray}
	\Phi^{\dagger}(x,\theta ,\bar{\theta}) \! \! \! \! \! \! \! 
	& & =e^{\frac{1}{2}\theta \sigma^{\mu}\bar{\theta}\Theta^{-1}_{\mu \nu}
	x^{\nu}} \{ A^{\dagger}(y^{\dagger})+\sqrt{2}\bar{\theta} \bar{\psi}
	(y^{\dagger})+\bar{\theta}\bar{\theta} F^{\dagger}(y^{\dagger})\},
	\quad  y^{\dagger \mu}=x^{\mu}-i\theta \sigma^{\mu}\bar{\theta} 
	\nonumber \\
	& & =A^{\dagger}(x)+\theta \sigma^{\mu}\bar{\theta}
	\hat{p}^{\prime}_{\mu}A^{\dagger}(x)-\frac{1}{4}\theta \theta 
	\bar{\theta}\bar{\theta}\hat{p}^{\prime}_{\mu}\hat{p}^{\prime \mu}
	A^{\dagger}(x) \nonumber \\
	& & \qquad \qquad 
	+\sqrt{2}\bar{\theta}\bar{\psi}(x)-\frac{1}{\sqrt{2}}
	\bar{\theta}\bar{\theta}\theta \sigma^{\mu}\hat{p}^{\prime}_{\mu}
	\bar{\psi}(x)+\bar{\theta}\bar{\theta} F^{\dagger}(x).
\end{eqnarray}

\noindent By making use of these fields, we can obtain the supersymmetric action in the following form: 

\begin{eqnarray}
	S \! \! \! \! \! \! \! 
	& & =\int d^4 x d\theta d\theta d\bar{\theta} d\bar{\theta} \> \> 
	e^{-\frac{1}{2}\theta \sigma^{\mu}\bar{\theta}\Theta^{-1}_{\mu \nu}
	x^{\nu}} \Phi^{\dagger}(x,\theta ,\bar{\theta}) 
	\Phi (x,\theta ,\bar{\theta}) \\
	& & =\int d^4 x \Big \{ -\hat{p}^{\prime}_{\mu}A^{\dagger}(x) 
	\hat{p}^{\mu}A(x)+\bar{\psi}(x)\bar{\sigma}^{\mu}\hat{p}_{\mu}\psi (x)
	+F^{\dagger}(x)F(x) \nonumber \\
	& & \qquad \qquad \qquad 
	+\big [ -\frac{1}{2}A^{\dagger}(x)\hat{p}^{\mu}
	\Theta^{-1}_{\mu \nu}x^{\nu}A(x)-\frac{1}{4}\psi (x)\sigma^{\mu}
	\Theta^{-1}_{\mu \nu}x^{\nu}\bar{\psi}(x) \nonumber \\
	& & \qquad \qquad \qquad \quad \> \; 
	-\frac{1}{\sqrt{6}} \eta^{\mu \nu}\Theta^{-1}_{\mu \rho}x^{\rho}
	\Theta^{-1}_{\nu \lambda}x^{\lambda}A^{\dagger}(x)A(x)\big ] \Big \}.
\end{eqnarray}

\noindent The $\big [ \cdots \big ]$ term in (33) is from the integration measure $e^{-\frac{1}{2}\theta \sigma^{\mu}\bar{\theta}\Theta^{-1}_{\mu \nu}x^{\nu}}$ which ensures the supersymmetry-invariance of the action. Besides this counter term, the bosonic part of (33) reproduces (4) for massless case. So, the representation (21)(22) seems fairly well.

\vskip 1.5cm

%%%%%%%%%%%%%%%%%%%%%%%%%%%%%%%%%%%%%%%%%%%%%%%%%%%%%%%%%%%%%%%%%%%%%%%%%%%%%%%
\noindent
{\bf 4. Supersymmetric gauge theory on noncommutative space}

\vskip 0.5cm

This section will be allotted for the supersymmetric gauge theory, and we will see that it has a close connection with IIB matrix model. Before turning to the task, I recommend you to see Appendices for the definition of the trace. In order to construct the gauge theory, we need the trace which has cyclic symmetry $\displaystyle \int ab=\int ba$ to ensure the gauge-invariance. However, for example, the representation (2)(7) is essentially the harmonic oscillator, i.e. infinite dimensional matrix, so the usual trace that sums up the diagonal components is not suitable for the purpose. After you see Appendices, the discussion in this section can be done, independent of the representation of $\mathcal{A}\oplus End_{\mathcal{A}}\mathcal{H}$. Now, we fix $\displaystyle \int$ denotes the trace that has cyclic symmetry.

As we have seen in \S 2, the gauge fields live in $\mathcal{A}$ and $End_{\mathcal{A}}\mathcal{H}$ respectively, so we must think of the superfields generated from $\mathcal{A}\oplus \{ \theta ,\bar{\theta}\}$ and $End_{\mathcal{A}}\mathcal{H}\oplus \{ \theta ,\bar{\theta}\}$. Because these superfields have the same forms as usual ones, we can expand them as follows: 

\begin{eqnarray}
	& & F(\hat{x},\theta ,\bar{\theta})=f(\hat{x})+\theta \phi (\hat{x})
	+\bar{\theta}\bar{\chi}(\hat{x})+\theta \theta m(\hat{x})
	+\bar{\theta}\bar{\theta}n(\hat{x}) \nonumber \\
	& & \qquad \qquad \qquad \qquad 
	+\theta \sigma^{\mu}\bar{\theta}v_{\mu}(\hat{x})
	+\theta \theta \bar{\theta}\bar{\lambda}(\hat{x})
	+\bar{\theta}\bar{\theta}\theta \psi (\hat{x})
	+\theta \theta \bar{\theta}\bar{\theta}d(\hat{x}), \\
	& & F(\hat{x}^{\prime},\theta ,\bar{\theta})=f(\hat{x}^{\prime})
	+\theta \phi (\hat{x}^{\prime})+\bar{\theta}\bar{\chi}(\hat{x}^{\prime})
	+\theta \theta m(\hat{x}^{\prime})+\bar{\theta}\bar{\theta}
	n(\hat{x}^{\prime}) \nonumber \\
	& & \qquad \qquad \qquad \qquad 
	+\theta \sigma^{\mu}\bar{\theta}v_{\mu}(\hat{x}^{\prime})
	+\theta \theta \bar{\theta}\bar{\lambda}(\hat{x}^{\prime})
	+\bar{\theta}\bar{\theta}\theta \psi (\hat{x}^{\prime})
	+\theta \theta \bar{\theta}\bar{\theta}d(\hat{x}^{\prime}).
\end{eqnarray}

For these superfields, let us define the supersymmetry transformations in the algebra $\mathcal{A}\oplus End_{\mathcal{A}}\mathcal{H}$.

\begin{eqnarray}
	& & \delta_{\xi}F(\hat{x},\theta ,\bar{\theta}) \equiv [\xi Q
	+\bar{\xi}\bar{Q},F(\hat{x},\theta ,\bar{\theta})] \\
	& & \delta_{\xi}F(\hat{x}^{\prime},\theta ,\bar{\theta}) \equiv 
	[\xi Q+\bar{\xi}\bar{Q},F(\hat{x}^{\prime},\theta ,\bar{\theta})]
\end{eqnarray}

\noindent Under these transformations, we can obtain the invariant action in $\mathcal{A}\oplus End_{\mathcal{A}}\mathcal{H}$ by using the integral $\displaystyle \int$.

\begin{eqnarray}
	& & I=\int \big (\int d\theta d\theta d\bar{\theta} d\bar{\theta} \> 
	F(\hat{x},\theta ,\bar{\theta})+\int d\theta d\theta d\bar{\theta} 
	d\bar{\theta} \> \> F(\hat{x}^{\prime},\theta ,\bar{\theta}) \big ) \\
	& & \delta_{\xi}I=\int \big ( \delta_{\xi}d(\hat{x})+\delta_{\xi}
	d(\hat{x}^{\prime})\big )=\int \big ( -\frac{1}{2}\xi \sigma^{\mu}
	[\hat{p}_{\mu},\bar{\lambda}(\hat{x})]-\frac{1}{2}\bar{\xi} 
	\bar{\sigma}^{\mu}[\hat{p}^{\prime}_{\mu},\psi (\hat{x})] \big )=0 
	\nonumber
\end{eqnarray}

Later, in order to use as the gauge transformation, let us define the chiral, anti-chiral superfields living in $\mathcal{A}$ and $End_{\mathcal{A}}\mathcal{H}$. Their definitions are 

\begin{eqnarray}
	& & \text{chiral superfield} \quad [\bar{D}_{\dot{\alpha}},\Phi ]=0, \\
	& & \text{anti-chiral superfield} \quad [D_{\alpha},\Phi^{\dagger}]=0.
\end{eqnarray}

\noindent Using the following relations, 

\begin{eqnarray*}
	& & [\bar{D}_{\dot{\alpha}},\hat{x}^{\mu}+i\theta \sigma^{\mu}
	\bar{\theta}]=[\bar{D}_{\dot{\alpha}},\hat{x}^{\prime \mu}]=0, \\
	& & [D_{\alpha},\hat{x}^{\mu}]=[D_{\alpha},\hat{x}^{\prime \mu}
	-i\theta \sigma^{\mu}\bar{\theta}]=0,
\end{eqnarray*}

\noindent we obtain the general solutions for (39)(40).

\begin{eqnarray}
	\Phi (\hat{x},\theta ,\bar{\theta}) \! \! \! \! \! \! \! \! 
	& & =A(\hat{x}+i\theta \sigma\bar{\theta})+\sqrt{2}\theta \psi (\hat{x}
	+i\theta \sigma\bar{\theta})+\theta \theta F(\hat{x}+i\theta \sigma 
	\bar{\theta}) \nonumber \\
	& & \equiv \Phi (\hat{x}+i\theta \sigma \bar{\theta}) \\
	\Phi (\hat{x}^{\prime},\theta ,\bar{\theta}) \! \! \! \! \! \! \! \! 
	& & =A(\hat{x}^{\prime})+\sqrt{2}\theta \psi (\hat{x}^{\prime})
	+\theta \theta F(\hat{x}^{\prime}) \nonumber \\
	& & \equiv \Phi (\hat{x}^{\prime}) \\
	\Phi^{\dagger}(\hat{x},\theta ,\bar{\theta}) \! \! \! \! \! \! \! \! 
	& & =A^{\dagger}(\hat{x})+\sqrt{2}\bar{\theta}\bar{\psi}(\hat{x})
	+\bar{\theta}\bar{\theta}F^{\dagger}(\hat{x}) \nonumber \\
	& & \equiv \Phi^{\dagger}(\hat{x}) \\
	\Phi^{\dagger}(\hat{x}^{\prime},\theta ,\bar{\theta}) \! \! \! \! \! 
	\! \! \! 
	& & =A^{\dagger}(\hat{x}^{\prime}-i\theta \sigma \bar{\theta})+\sqrt{2}
	\bar{\theta}\bar{\psi}(\hat{x}^{\prime}-i\theta \sigma \bar{\theta})
	+\bar{\theta}\bar{\theta}F^{\dagger}(\hat{x}^{\prime}-i\theta \sigma 
	\bar{\theta}) \nonumber \\
	& & \equiv \Phi^{\dagger}(\hat{x}^{\prime}-i\theta \sigma \bar{\theta})
\end{eqnarray}

\noindent The supersymmetric integral for chiral, anti-chiral superfields can be obtained as 

\begin{eqnarray}
	I_c=\int \big (\int d\theta d\theta \Big \{ \Phi_1(\hat{x},\theta 
	,\bar{\theta})+\Phi_2(\hat{x}^{\prime},\theta ,\bar{\theta})\Big \} 
	+\int d\bar{\theta}d\bar{\theta} \Big \{ \Phi^{\dagger}_1(\hat{x}
	,\theta ,\bar{\theta})+\Phi^{\dagger}_2(\hat{x}^{\prime},\theta 
	,\bar{\theta}) \Big \} \big ).
\end{eqnarray}

\noindent $\delta_{\xi}I_c=0$ owes to the cyclic symmetry of the trace.

At last, let us introduce the gauge fields as the vector superfields. These are defined in a usual way and expanded as 

\begin{eqnarray}
	& & V(\hat{x},\theta ,\bar{\theta})=V^{\dagger}(\hat{x},\theta 
	,\bar{\theta}), \nonumber \\
	& & V(\hat{x},\theta ,\bar{\theta})=C(\hat{x})+i\theta \chi (\hat{x})
	-i\bar{\theta}\bar{\chi}(\hat{x}) \nonumber \\
	& & \qquad \qquad \qquad 
	+\frac{i}{2}\theta \theta \big \{ M(\hat{x})+iN(\hat{x})\big \} 
	-\frac{i}{2}\bar{\theta}\bar{\theta}\big \{ M(\hat{x})-iN(\hat{x})
	\big \} -\theta \sigma^{\mu}\bar{\theta}v_{\mu}(\hat{x}) \nonumber \\
	& & \qquad \qquad \qquad 
	+i\theta \theta \bar{\theta}\big \{ \bar{\lambda}(\hat{x})-\frac{1}{4}
	\bar{\sigma}^{\mu}[\hat{p}_{\mu},\chi(\hat{x})]\big \}-i\bar{\theta}
	\bar{\theta}\theta \big \{ \lambda (\hat{x})-\frac{1}{4}\sigma^{\mu}
	[\hat{p}_{\mu},\bar{\chi}(\hat{x})]\big \} \nonumber \\
	& & \qquad \qquad \qquad 
	+\frac{1}{2}\theta \theta \bar{\theta}\bar{\theta}\big \{ D(\hat{x})
	-\frac{1}{8}[\hat{p}_{\mu},[\hat{p}^{\mu},C(\hat{x})]]\big \} , \\
	& & \nonumber \\
	& & V(\hat{x}^{\prime},\theta ,\bar{\theta})=V^{\dagger}
	(\hat{x}^{\prime},\theta ,\bar{\theta}), \nonumber \\
	& & V(\hat{x}^{\prime},\theta ,\bar{\theta})=C(\hat{x}^{\prime})
	+i\theta \chi (\hat{x}^{\prime})-i\bar{\theta}\bar{\chi}
	(\hat{x}^{\prime}) \nonumber \\
	& & \qquad \qquad \qquad 
	+\frac{i}{2}\theta \theta \big \{ M(\hat{x}^{\prime})
	+iN(\hat{x}^{\prime})\big \} -\frac{i}{2}\bar{\theta}\bar{\theta}
	\big \{ M(\hat{x}^{\prime})-iN(\hat{x}^{\prime})\big \} 
	-\theta \sigma^{\mu}\bar{\theta}v_{\mu}(\hat{x}^{\prime}) \nonumber \\
	& & \qquad \qquad \qquad 
	+i\theta \theta \bar{\theta}\big \{ \bar{\lambda}(\hat{x}^{\prime})
	-\frac{1}{4}\bar{\sigma}^{\mu}[\hat{p}^{\prime}_{\mu}
	,\chi(\hat{x}^{\prime})]\big \}-i\bar{\theta}\bar{\theta}\theta 
	\big \{ \lambda (\hat{x}^{\prime})-\frac{1}{4}\sigma^{\mu}
	[\hat{p}^{\prime}_{\mu},\bar{\chi}(\hat{x}^{\prime})]\big \} 
	\nonumber \\
	& & \qquad \qquad \qquad 
	+\frac{1}{2}\theta \theta \bar{\theta}\bar{\theta}\big \{ 
	D(\hat{x}^{\prime})-\frac{1}{8}[\hat{p}^{\prime}_{\mu}
	,[\hat{p}^{\prime \mu},C(\hat{x}^{\prime})]]\big \} .
\end{eqnarray}

Though the gauge transformation is realized $V\! \to \! V+\Phi +\Phi^{\dagger}+\cdots$ as we will see later, (41)(42)(43)(44) are not clear in chirality because of their coordinates. So, we define new generators.

\begin{eqnarray}
	\left \{ \begin{array}{l}
	\hat{X}^{\mu} \equiv \hat{x}^{\mu}+\displaystyle \frac{i}{2}\theta 
	\sigma^{\mu}\bar{\theta} \\
	\hat{X}^{\prime \mu} \equiv \hat{x}^{\prime \mu}-\displaystyle 
	\frac{\mathstrut i}{2}\theta \sigma^{\mu}\bar{\theta}
	\end{array} \right.
\end{eqnarray}

\noindent They satisfy the expected relations similar to (2)(7).

\begin{eqnarray}
	\left \{ \begin{array}{l}
	\hat{P}^{\mu}=-\Theta^{-1}_{\mu \nu}\hat{X}^{\nu} \\
	\hat{P}^{\prime \mu}=\Theta^{-1}_{\mu \nu}\hat{X}^{\prime \nu}
	\end{array} \right.
\end{eqnarray}

Then, the gauge transformation becomes 

\begin{eqnarray*}
	& & V(\hat{X}) \to V(\hat{X})+\Phi (\hat{X}+\frac{i}{2}\theta 
	\sigma^{\mu}\bar{\theta})+\Phi^{\dagger}(\hat{X}-\frac{i}{2}\theta 
	\sigma^{\mu}\bar{\theta})+\cdots , \\
	& & V(\hat{X}^{\prime}) \to V(\hat{X}^{\prime})+\Phi (\hat{X}^{\prime}
	+\frac{i}{2}\theta \sigma^{\mu}\bar{\theta})+\Phi^{\dagger}
	(\hat{X}^{\prime}-\frac{i}{2}\theta \sigma^{\mu}\bar{\theta})+\cdots .
	 \\
	& & \\
	& & \Phi (\hat{X}+\frac{i}{2}\theta \sigma^{\mu}\bar{\theta})
	+\Phi^{\dagger}(\hat{X}-\frac{i}{2}\theta \sigma^{\mu}\bar{\theta}) \\
	& & \qquad \qquad 
	=A(\hat{X})+A^{\dagger}(\hat{X})+\sqrt{2}\big \{ \theta \psi (\hat{X})
	+\bar{\theta}\bar{\psi}(\hat{X})\big \} +\theta \theta F(\hat{X})
	+\bar{\theta}\bar{\theta}F^{\dagger}(\hat{X}) \\
	& & \qquad \qquad \qquad 
	-\frac{1}{2}\theta \sigma^{\mu}\bar{\theta}[\hat{P}_{\mu},A(\hat{X})
	-A^{\dagger}(\hat{X})] \\
	& & \qquad \qquad \qquad 
	-\frac{1}{2\sqrt{2}}\theta \theta \bar{\theta}\bar{\sigma}^{\mu}
	[\hat{P}_{\mu},\psi (\hat{X})]-\frac{1}{2\sqrt{2}}\bar{\theta}
	\bar{\theta}\theta \sigma^{\mu}[\hat{P}_{\mu},\bar{\psi}(\hat{X})] \\
	& & \qquad \qquad \qquad 
	-\frac{1}{16}\theta \theta \bar{\theta}\bar{\theta}[\hat{P}_{\mu},
	[\hat{P}^{\mu},A(\hat{X})+A^{\dagger}(\hat{X})]] \\
	& & \Phi (\hat{X}^{\prime}+\frac{i}{2}\theta \sigma^{\mu}\bar{\theta})
	+\Phi^{\dagger}(\hat{X}^{\prime}-\frac{i}{2}\theta \sigma^{\mu}
	\bar{\theta}) \\
	& & \qquad \qquad 
	=A(\hat{X}^{\prime})+A^{\dagger}(\hat{X}^{\prime})+\sqrt{2}\big \{ 
	\theta \psi (\hat{X}^{\prime})+\bar{\theta}\bar{\psi}(\hat{X}^{\prime})
	\big \} +\theta \theta F(\hat{X}^{\prime})+\bar{\theta}\bar{\theta}
	F^{\dagger}(\hat{X}^{\prime}) \\
	& & \qquad \qquad \qquad 
	-\frac{1}{2}\theta \sigma^{\mu}\bar{\theta}[\hat{P}^{\prime}_{\mu}
	,A(\hat{X}^{\prime})-A^{\dagger}(\hat{X}^{\prime})] \\
	& & \qquad \qquad \qquad 
	-\frac{1}{2\sqrt{2}}\theta \theta \bar{\theta}\bar{\sigma}^{\mu}
	[\hat{P}^{\prime}_{\mu},\psi (\hat{X}^{\prime})]-\frac{1}{2\sqrt{2}}
	\bar{\theta}\bar{\theta}\theta \sigma^{\mu}[\hat{P}^{\prime}_{\mu}
	,\bar{\psi}(\hat{X}^{\prime})] \\
	& & \qquad \qquad \qquad 
	-\frac{1}{16}\theta \theta \bar{\theta}\bar{\theta}
	[\hat{P}^{\prime}_{\mu},[\hat{P}^{\prime \mu},A(\hat{X})+A^{\dagger}
	(\hat{X}^{\prime})]]
\end{eqnarray*}

\noindent Comparing them with (46)(47), we see that we can take the special gauge known as WZ gauge such as $C=\chi =M=N=0$.

Next, let us define the supersymmetric field strengths living in $\mathcal{A}$ and $End_{\mathcal{A}}\mathcal{H}$ respectively. 

\begin{eqnarray}
	& & \bar{W}_{\dot{\alpha}}(\hat{X}) \equiv +\frac{1}{4}[D,\{ D
	,e^{V(\hat{X})}[\bar{D}_{\dot{\alpha}},e^{-V(\hat{X})}]\} ],
	\quad \{ D_{\alpha},\bar{W}_{\dot{\alpha}}\}=0 \quad 
	\text{(anti-chiral)} \\
	& & W_{\alpha}(\hat{X}^{\prime}) \equiv -\frac{1}{4}[\bar{D},\{ \bar{D}
	,e^{-V(\hat{X}^{\prime})}[D_{\alpha},e^{V(\hat{X}^{\prime})}]\} ],
	\quad \{ \bar{D}_{\dot{\alpha}},W_{\alpha}\}=0 \quad \text{(chiral)}
\end{eqnarray}

\noindent Under the gauge transformation 

\begin{eqnarray*}
	& & e^{V(\hat{X})} \to e^{-i\Lambda^{\dagger}(\hat{X})}e^{V(\hat{X})}
	e^{i\Lambda (\hat{X})}, \\
	& & e^{V(\hat{X}^{\prime})} \to e^{-i\Lambda^{\dagger}
	(\hat{X}^{\prime})}e^{V(\hat{X}^{\prime})}e^{i\Lambda 
	(\hat{X}^{\prime})},
\end{eqnarray*}

\noindent they transform as 

\begin{eqnarray}
	& & \bar{W}_{\dot{\alpha}}(\hat{X}) \to e^{-i\Lambda^{\dagger}
	(\hat{X})}\bar{W}_{\dot{\alpha}}(\hat{X})e^{i\Lambda^{\dagger}
	(\hat{X})}, \\
	& & W_{\alpha}(\hat{X}^{\prime}) \to e^{-i\Lambda(\hat{X}^{\prime})}
	W_{\alpha}(\hat{X}^{\prime})e^{i\Lambda	(\hat{X}^{\prime})}.
\end{eqnarray}

Finally, we give the supersymmetic Yang-Mills action.

\begin{eqnarray}
	S_{YM}=-\int \big ( \int d\theta d\theta \> W^{\alpha}
	(\hat{X}^{\prime}) W_{\alpha}(\hat{X}^{\prime})
	+\int d\bar{\theta}d\bar{\theta} \> \bar{W}_{\dot{\alpha}}(\hat{X})
	\bar{W}^{\dot{\alpha}}(\hat{X}) \big )
\end{eqnarray}

Let us write down their concrete form in WZ gauge. Then, 

\begin{eqnarray*}
	& & V(\hat{X})=-\theta \sigma^{\mu}\bar{\theta}v_{\mu}(\hat{x})
	+i\theta \theta \bar{\theta}\bar{\lambda}(\hat{x})-i\bar{\theta}
	\bar{\theta}\theta \lambda(\hat{x})+\frac{1}{2}\theta \theta 
	\bar{\theta}\bar{\theta}\big (D(\hat{x})+\frac{1}{2}[\hat{p}_{\mu}
	,v^{\mu}(\hat{x})]\big ), \\
	& & V^2(\hat{X})=-\frac{1}{2}\theta \theta \bar{\theta}\bar{\theta}
	v_{\mu}(\hat{x})v^{\mu}(\hat{x}), \\
	& & V^3(\hat{X})=0, \\
	& & \\
	& & V(\hat{X}^{\prime})=-\theta \sigma^{\mu}\bar{\theta}v_{\mu}
	(\hat{x}^{\prime})+i\theta \theta \bar{\theta}\bar{\lambda}
	(\hat{x}^{\prime})-i\bar{\theta}\bar{\theta}\theta \lambda
	(\hat{x}^{\prime})+\frac{1}{2}\theta \theta \bar{\theta}\bar{\theta}
	\big (D(\hat{x}^{\prime})+\frac{1}{2}[\hat{p}^{\prime}_{\mu},v^{\mu}
	(\hat{x}^{\prime})]\big ), \\
	& & V^2(\hat{X}^{\prime})=-\frac{1}{2}\theta \theta \bar{\theta}
	\bar{\theta}v_{\mu}(\hat{x}^{\prime})v^{\mu}(\hat{x}^{\prime}), \\
	& & V^3(\hat{X}^{\prime})=0.
\end{eqnarray*}

\noindent Therefore the field strengths are 

\begin{eqnarray}
	\bar{W}_{\dot{\alpha}}(\hat{X}) \! \! \! \! \! \! \! 
	& & =i\bar{\lambda}_{\dot{\alpha}}
	(\hat{x})+\bar{\theta}_{\dot{\alpha}}D(\hat{x}) \nonumber \\
	& & \qquad -\frac{1}{4}\bar{\theta}^{\dot{\beta}}
	\epsilon_{\dot{\alpha}\dot{\gamma}}(\bar{\sigma}^{\mu}
	\sigma^{\nu})^{\dot{\gamma}}_{\> \> \dot{\beta}}\big \{ [\hat{p}_{\mu}
	,v_{\nu}(\hat{x})]-[\hat{p}_{\nu},v_{\mu}(\hat{x})]+[v_{\mu}(\hat{x})
	,v_{\nu}(\hat{x})]\big \} \nonumber \\
	& & \qquad 
	-\frac{i}{2}\bar{\theta}\bar{\theta}\epsilon_{\dot{\alpha}\dot{\beta}}
	\bar{\sigma}^{\mu \dot{\beta}\alpha}[\hat{p}_{\mu}+v_{\mu}(\hat{x})
	,\lambda_{\alpha}(\hat{x})] \nonumber \\
	& & =i\bar{\lambda}_{\dot{\alpha}}
	(\hat{x})+\bar{\theta}_{\dot{\alpha}}D(\hat{x})-\frac{1}{4}
	\bar{\theta}^{\dot{\beta}}\epsilon_{\dot{\alpha}\dot{\gamma}}
	(\bar{\sigma}^{\mu}\sigma^{\nu})^{\dot{\gamma}}_{\> \> \dot{\beta}}
	\big \{ [A_{\mu}(\hat{x}),A_{\nu}(\hat{x})]+i\Theta^{-1}_{\mu \nu}
	\big \} \nonumber \\
	& & \qquad 
	-\frac{i}{2}\bar{\theta}\bar{\theta}\epsilon_{\dot{\alpha}\dot{\beta}}
	\bar{\sigma}^{\mu \dot{\beta}\alpha}[A_{\mu}(\hat{x}),\lambda_{\alpha}
	(\hat{x})], \\
	& & \nonumber \\
	W_{\alpha}(\hat{X}^{\prime}) \! \! \! \! \! \! \! 
	& & =-i\lambda_{\alpha}
	(\hat{x}^{\prime})+\theta_{\alpha}D(\hat{x}^{\prime}) \nonumber \\
	& & \qquad +\frac{1}{4}\theta_{\beta}(\bar{\sigma}^{\mu}
	\sigma^{\nu})^{\> \> \beta}_{\alpha}\big \{ [\hat{p}^{\prime}_{\mu}
	,v_{\nu}(\hat{x}^{\prime})]-[\hat{p}^{\prime}_{\nu},v_{\mu}
	(\hat{x}^{\prime})]+[v_{\mu}(\hat{x}^{\prime}),v_{\nu}
	(\hat{x}^{\prime})]\big \} \nonumber \\
	& & \qquad 
	+\frac{i}{2}\theta \theta \sigma^{\mu}_{\alpha \dot{\alpha}}
	[\hat{p}^{\prime}_{\mu}+v_{\mu}(\hat{x}^{\prime})
	,\lambda^{\dot{\alpha}}(\hat{x}^{\prime})] \nonumber \\
	& & =-i\lambda_{\alpha}
	(\hat{x}^{\prime})+\theta_{\alpha}D(\hat{x}^{\prime})+\frac{1}{4}
	\theta_{\beta}(\sigma^{\mu}\bar{\sigma}^{\nu})^{\> \> \beta}_{\alpha}
	\big \{ [A_{\mu}(\hat{x}^{\prime}),A_{\nu}(\hat{x}^{\prime})]
	-i\Theta^{-1}_{\mu \nu}\big \} \nonumber \\
	& & \qquad 
	+\frac{i}{2}\theta \theta \sigma^{\mu}_{\alpha \dot{\alpha}}[A_{\mu}
	(\hat{x}^{\prime}),\bar{\lambda}^{\dot{\alpha}}(\hat{x}^{\prime})].
\end{eqnarray}

\noindent Here, we have defined as $A_{\mu}(\hat{x})\equiv \hat{p}_{\mu}+v_{\mu}(\hat{x})$ and $A_{\mu}(\hat{x}^{\prime})\equiv \hat{p}^{\prime}_{\mu}+A_{\mu}(\hat{x}^{\prime})$. In this gauge, the supersymmetric Yang-Mills action becomes 

\begin{eqnarray}
	S_{YM} \! \! \! \! \! \! \! 
	& & =-\int \big ( \int d\theta d\theta \> W^{\alpha}
	(\hat{X}^{\prime}) W_{\alpha}(\hat{X}^{\prime})
	+\int d\bar{\theta}d\bar{\theta} \> \bar{W}_{\dot{\alpha}}(\hat{X})
	\bar{W}^{\dot{\alpha}}(\hat{X}) \big ) \nonumber \\
	& & =-\int \big ( \frac{1}{8}[A_{\mu}(\hat{x}^{\prime})
	,A_{\nu}(\hat{x}^{\prime})][A^{\mu}(\hat{x}^{\prime}),A^{\nu}
	(\hat{x}^{\prime})]+\lambda (\hat{x}^{\prime})\sigma^{\mu}
	[A_{\mu}(\hat{x}^{\prime}),\bar{\lambda}(\hat{x}^{\prime})] 
	\nonumber \\
	& & \qquad \qquad \qquad \qquad \qquad \qquad \qquad \qquad \qquad 
	+D(\hat{x}^{\prime})D(\hat{x}^{\prime})
	-\frac{1}{8}\Theta^{-1}_{\mu \nu}\Theta^{-1\mu \nu}\big ) \nonumber \\
	& & \quad -\int \big ( \frac{1}{8}[A_{\mu}(\hat{x}),A_{\nu}(\hat{x})]
	[A^{\mu}(\hat{x}),A^{\nu}(\hat{x})]+\lambda (\hat{x})\sigma^{\mu}
	[A_{\mu}(\hat{x}),\bar{\lambda}(\hat{x})] \nonumber \\
	& & \qquad \qquad \qquad \qquad \qquad \qquad \qquad \qquad \qquad 
	+D(\hat{x})D(\hat{x})
	-\frac{1}{8}\Theta^{-1}_{\mu \nu}\Theta^{-1\mu \nu}\big ) \nonumber \\
	& & =-\int \big ( \frac{1}{4}[A_{\mu}(\hat{x}),A_{\nu}(\hat{x})]
	[A^{\mu}(\hat{x}),A^{\nu}(\hat{x})]-2\bar{\lambda}(\hat{x})
	\bar{\sigma}^{\mu}[A_{\mu}(\hat{x}),\lambda (\hat{x})] \nonumber \\
	& & \qquad \qquad \qquad \qquad \qquad \qquad \qquad \qquad \qquad 
	+2D(\hat{x})D(\hat{x})
	-\frac{1}{4}\Theta^{-1}_{\mu \nu}\Theta^{-1\mu \nu}\big ).
\end{eqnarray}

\noindent The auxiliary field $D(\hat{x})$ can be integrated out, so we omit it in the following. 

This is the supersymmetric Yang-Mills action on noncommutative space for any representation of $\mathcal{A}\oplus End_{\mathcal{A}}\mathcal{H}$. This looks very like IIB matrix model in its form. In fact, if we take $d=10$, i.e. $\mu ,\nu ,\cdots$ be SO(10) vector indices and $\alpha ,\beta ,\cdots$ SO(10) Weyl spinor indices, then, 

\begin{eqnarray}
	S_{YM}=-\int \! \big ( \frac{1}{4}[A_{\mu}(\hat{x}),A_{\nu}(\hat{x})]
	[A^{\mu}(\hat{x}),A^{\nu}(\hat{x})]-\frac{1}{2}\bar{\psi}(\hat{x})
	\Gamma^{\mu}[A_{\mu}(\hat{x}),\psi (\hat{x})]-\frac{1}{4}
	\Theta^{-1}_{\mu \nu}\Theta^{-1\mu \nu}\big ).
\end{eqnarray}

\noindent This action is just the matrix-regularized Green-Schwarz IIB string action in Schild gauge. Then, we can evaluate the parameter of noncommutativity $\Theta$ using the string tenson $T=\displaystyle \frac{1}{2\pi \alpha^{\prime}}$: 

\begin{eqnarray}
	T \sim \sqrt{\Theta^{-1}_{\mu \nu}\Theta^{-1\mu \nu}} \sim \Theta^{-1}
	, \qquad \alpha^{\prime} \sim \Theta
\end{eqnarray}

\noindent Independent of the representation of $\mathcal{A}\oplus End_{\mathcal{A}}\mathcal{H}$, the fact that the space-time noncommutativity must be at the same order as the string length is, in a sense, a desired result. That is to say, the space-time noncommutativity emerges at the scale the gravity is dominant. Therefore, the next aim will be the supergravity model on noncommutative space by using the representation (21).

\vskip 1.5cm

%%%%%%%%%%%%%%%%%%%%%%%%%%%%%%%%%%%%%%%%%%%%%%%%%%%%%%%%%%%%%%%%%%%%%%%%%%%%%%%
\noindent
{\bf 5. Discussions and acknowledgements}

\vskip 0.5cm

In this paper, we introduced a supersymmetry algebra and constructed supersymmetric field theories on noncommutative space. As a result, we saw that the scalar field theory constructed by means of chiral superfields agreed with that under the argument the first quantization is equivalent to making the space-time noncommutative, and the supersymmetric Yang-Mills theory on noncommutative space had a close relationship with IIB matrix model. There, the parameter $\Theta$ which expresses the space-time noncommutativity had the same order as the string scale $\alpha^{\prime}$. Especially, the last result is preferable because it shows the space-time noncommutativity emerges at the scale the gravity is dominant. Originally, the aim of introducing the space-time noncommutativity is to regularize the divergence arises from the short distance in the gravitational interaction. And, all through this paper, taking the naive limit $\Theta \! \to \! 0$ reproduces the ordinary field theories on commutative (geometric) space. So, it is reasonable to think that the divergence-free, consistent quantum gravity model can be obtained on noncommutative (algebraic) space. And, with the supersymmetry algebra on noncommutative space, we can expect to construct a supergravity model in a obvious way. Therefore, as an very interesting future work, the supergravity model with the representation (21)(22) will be appeared soon.

\vskip 0.5cm

\noindent \underline{ Acknowledgements }

\vskip 0.2cm

I would like to thank my colleagues for useful comments and discussions. Especially, I gratefully acknowledge S. Shinohara for helpful suggestions on the supersymmetry-invariance of scalar field theory.

%%%%%%%%%%%%%%%%%%%%%%%%%%%%%%%%%%%%%%%%%%%%%%%%%%%%%%%%%%%%%%%%%%%%%%%%%%%%%%%
\newpage
\noindent
{\bf Appendix 1.}

\noindent
{\bf Trace with cyclic symmetry under the representation (2)(7)}

\vskip 0.5cm

First, we present the spectral triple $(\mathcal{A}\oplus End_{\mathcal{A}}\mathcal{H},\mathcal{H},D)$. The representation of $\mathcal{A}\oplus End_{\mathcal{A}}\mathcal{H}$ is 

\begin{align}
	& \left. \begin{array}{l}
	\hat{x}^{\mu}=\displaystyle \frac{1}{2}x^{\mu}+i\Theta^{\mu \nu}
	\partial_{\nu}, \quad \hat{p}_{\mu}=-\Theta^{-1}_{\mu \nu}
	\hat{x}^{\nu}, \\
	\hat{x}^{\prime \mu}=\displaystyle \frac{\mathstrut 1}{2}x^{\mu}
	-i\Theta^{\mu \nu}\partial_{\nu}, \quad \hat{p}^{\prime}_{\mu}
	=\Theta^{-1}_{\mu \nu}\hat{x}^{\prime \nu}. 
	\end{array} \right. \tag{A.1} \\
	& \qquad \mu ,\nu=0,\cdots ,(d-1) \notag
\end{align}

\noindent $\mathcal{H}$ is the space of square-integrable functions on $\mathbf{R}^{d-1,1}$. ``Dirac operator" $D$ is defined as 

\begin{align}
	D & \equiv \frac{1}{2}(\hat{p}_{\mu}\hat{p}^{\mu}
	+\hat{p}^{\prime}_{\mu}\hat{p}^{\prime \mu}) \tag{A.2} \\
	& =-\partial_{\mu}\partial^{\mu}+\frac{1}{4}\eta^{\mu \nu}
	\Theta^{-1}_{\mu \rho}x^{\rho}\Theta^{-1}_{\nu \lambda}x^{\lambda}. 
	\notag
\end{align}

\noindent For simplicity, we set $\big |\Theta^{-1}_{\mu k}\Theta^{-1}_{\nu k}\big |=\Theta^{-2}, \> (k=0,\cdots ,(d-1))$. Essentially, $|D|\equiv (D^{\dagger}D)^{\frac{1}{2}}$ is the $d$-dimensional harmonic oscillator whose oscillation number is $\Theta^{-1}$. Therefore, its spectrum becomes 

\begin{align*}
	& \text{eigenvalue} \qquad \sum_{\mu =0}^{d-1}\frac{1}{\Theta} (n_{\mu}
	+\frac{1}{2})=\frac{1}{\Theta}(N+\frac{d}{2})\equiv \mu_{N}(|D|), \\
	& \text{multiplicity} \qquad {}_{N+d-1}C_{d-1}
	=\textstyle \frac{(N+d-1)!}{N!(d-1)!}\equiv m_N.
\end{align*}

\noindent As we see from this spectrum, the ordinary trace is not appropriate for our purpose, because that of infinite-dimensional matrix does not have cyclic symmetry, $tr(ab)\neq tr(ba)$. So we must introduce a special trace known as Dixmier trace, which, so to speak, draws out the coefficient of the logarithmic divergence. In practice, let us calculate Dixmier trace $tr_{\omega}(|D|^s)$.

\begin{align}
	tr_{\omega}(|D|^{-d}) & =\lim_{N\to \infty}
	\frac{1}{\Bigg (\log \displaystyle \sum_{n=0}^{N}m_n\Bigg )} 
	\sum_{n=0}^{N}m_n\mu_n(|D|^{-d}) \notag \\
	& \sim \lim_{N\to \infty}\frac{1}{d\log N}\Theta^{d}\log N \notag \\
	& \sim \frac{1}{d}\Theta^{d} \tag{A.3}
\end{align}

\begin{equation*}
	tr_{\omega}(|D|^s) 
	=\left \{ \begin{array}{l}
	\infty, \quad \text{for} \quad s>-d \\
	0, \quad \text{for} \quad s<-d
	\end{array} \right.
\end{equation*}

With this trace for $s=-d$, we can define the regularized integral with cyclic symmetry as follows.

\begin{align}
	& \int a \equiv d\Theta^{-d} \> tr_{\omega} (|D|^{-d} a), \qquad a\in 
	\mathcal{A}\oplus End_{\mathcal{A}}\mathcal{H} \tag{A.4} \\
	& \qquad \int ab=\int ba, \quad \int 1=1 \notag
\end{align}

\noindent $|D|^{-d}$ is (two-sided) ideal which transfers $\forall a\in \mathcal{A}\oplus End_{\mathcal{A}}\mathcal{H}$ to the trace class $\mathcal{L}^{(1,\infty)}$, that is to say, makes the trace finite. Due to this Dixmier trace, we can obtain the gauge-invariance of Yang-Mills theory on noncommutative space.

\vskip 1.5cm

%%%%%%%%%%%%%%%%%%%%%%%%%%%%%%%%%%%%%%%%%%%%%%%%%%%%%%%%%%%%%%%%%%%%%%%%%%%%%%%
\noindent
{\bf Appendix 2.}

\noindent
{\bf Twist-eating solution ($SU(N)$ matrix representation of $\mathcal{A}\oplus End_{\mathcal{A}}\mathcal{H}$)}

\vskip 0.5cm

The representation (A.1) of the algebra $\mathcal{A}\oplus End_{\mathcal{A}}\mathcal{H}$ is the representation by the infinite-dimensional matrices. On the other hands, we can represent it by $SU(N)$ matrices. In this appendix, we review its representation known as twist-eating solution.

As the starting point, we write down the algabra $\mathcal{A}\oplus End_{\mathcal{A}}\mathcal{H}$ in the following form.

\begin{align}
	& U_{\mu}U_{\nu}=e^{i\Theta_{\mu \nu}}U_{\nu}U_{\mu} \notag \\
	& U^{\prime}_{\mu}U^{\prime}_{\nu}=e^{-i\Theta_{\mu \nu}}
	U^{\prime}_{\nu}U^{\prime}_{\mu} \tag{A.5} \\
	& U_{\mu}U^{\prime}_{\nu}=U^{\prime}_{\nu}U_{\mu} \notag \\
	& \qquad U_{\mu}=e^{\hat{x}^{\mu}}, \> U^{\prime}_{\mu}=
	e^{\hat{x}^{\prime \mu}}, \quad \mu ,\nu =1,\cdots ,d \notag
\end{align}

\noindent $\hat{x}_{\mu},\> \hat{x}^{\prime}_{\mu}$ are defined as 

\begin{align*}
	& \hat{x}_{\mu}=\log U_{\mu}\equiv \sum_{n=1}^{\infty} 
	\frac{(-1)^{n-1}}{n}(U_{\mu}-1)^n, \\
	& \hat{x}^{\prime}_{\mu}=\log U^{\prime}_{\mu}\equiv 
	\sum_{n=1}^{\infty} \frac{(-1)^{n-1}}{n}(U^{\prime}_{\mu}-1)^n.
\end{align*}

\noindent Any element is expanded for various $\mathcal{A}\oplus End_{\mathcal{A}}\mathcal{H}$ as 

\begin{align*}
	& a=\sum_{k_{\mu},k^{\prime}_{\mu}\in \mathbf{Z}}a(k,k^{\prime})
	e^{k_{\mu}\hat{x}^{\mu}+k^{\prime}_{\mu}\hat{x}^{\prime \mu}}, \qquad 
	\text{for noncommutative $d$-torus,} \\
	& a=\int d^d k d^d k^{\prime}\> a(k,k^{\prime})e^{k_{\mu}\hat{x}^{\mu}
	+k^{\prime}_{\mu}\hat{x}^{\prime \mu}}, \qquad 
	\text{for noncommutative $\mathbf{R}^d$.}
\end{align*}

\noindent Here, for special case, we take $d=2g$.

First, we write $\Theta_{\mu \nu}=\frac{2\pi}{N}M_{\mu \nu}$, and make $M_{\mu \nu}$ block-diagonal by making use of $SL(d,\mathbf{Z})$ symmetry in the following form.

\begin{equation}
	M=\left ( \begin{array}{cccc}
	M_1 \epsilon & & & \\
	 & M_2 \epsilon & & \\
	 & & \ddots & \\
	 & & & M_g \epsilon 
	\end{array} \right ), 
	\qquad \epsilon =\left ( \begin{array}{cc}
	0 & 1 \\ -1 & 0 \end{array} \right ) \tag{A.6}
\end{equation}

\noindent Then, the algebra $\mathcal{A}\oplus End_{\mathcal{A}}\mathcal{H}$ becomes 

\begin{align}
	& V_{\mu}V_{\nu}=e^{2\pi iM_{\mu \nu}/N}V_{\nu}V_{\mu}, \notag \\
	& V^{\prime}_{\mu}V^{\prime}_{\nu}=e^{-2\pi iM_{\mu \nu}/N}
	V^{\prime}_{\nu}V^{\prime}_{\mu}, \tag{A.7} \\
	& V_{\mu}V^{\prime}_{\nu}=V^{\prime}_{\nu}V_{\mu}. \notag
\end{align}

Next, we have $n_{\mu}$ be the greatest common divisor of $M_{\mu}$ and $N$: $n_{\mu}=\gcd (M_{\mu},N)$, and take $N_{\mu}\equiv \frac{N}{n_{\mu}}$. Now, $\mu =1,\cdots ,g$. Then, the twist-eating solution can exist only if $N_1\cdots N_g|N$. So we set $N=(N_1\cdots N_g)^2 N_0$. When this condition is satisfied, the representation is given by the tensor products of $SU(N_{\mu})$ matrices.

\begin{equation}
	\left. \begin{array}{l}
	V_{2\mu -1}=1_{N_1}\otimes \cdots \otimes P_{N_{\mu}} \otimes \cdots 
	\otimes 1_{N_g} \otimes 1_{N_1} \otimes \cdots \otimes 1_{N_g} 
	\otimes 1_{N_0} \\
	V_{2\mu}=1_{N_1}\otimes \cdots \otimes 
	(Q_{N_{\mu}})^{M_{\mu}/n_{\mu}} \otimes \cdots 
	\otimes 1_{N_g} \otimes 1_{N_1} \otimes \cdots \otimes 1_{N_g} 
	\otimes 1_{N_0} \\
	\\
	V^{\prime}_{2\mu -1}=1_{N_1} \otimes \cdots \otimes 1_{N_g} 
	\otimes 1_{N_1}\otimes \cdots \otimes P^{*}_{N_{\mu}} \otimes \cdots 
	\otimes 1_{N_g} \otimes 1_{N_0} \\
	V^{\prime}_{2\mu}=1_{N_1} \otimes \cdots \otimes 1_{N_g} 
	\otimes 1_{N_1}\otimes \cdots \otimes 
	(Q^{*}_{N_{\mu}})^{M_{\mu}/n_{\mu}} \otimes \cdots 
	\otimes 1_{N_g} \otimes 1_{N_0}
	\end{array} \right. \tag{A.8}
\end{equation}

\noindent Matrices $P_N,\> Q_N$ are $SU(N)$ clock and shift matrices and $P^{*}_N,\> Q^{*}_N$ are their complex conjugates.

\begin{equation}
	P_N=\left( \begin{array}{ccccc}
	0 & 1 &  & & \\
	& 0 & 1 & & \\
	& & \ddots & & \\
	& & & \ddots & 1 \\
	1 & & & & 0 \end{array} \right ),\quad 
	Q_N=\left ( \begin{array}{ccccc}
	1 & & & & \\
	& e^{2\pi i/N} & & & \\
	& & e^{4\pi i/N} & & \\
	& & & \ddots & \\
	& & & & e^{2\pi i(N-1)/N} \end{array} \right ) \tag{A.9}
\end{equation}

From these representative matrices, it is clear for ordinary trace that 

\begin{equation}
	tr(U_1^{k_1}\cdots U_d^{k_d}\> U_1^{\prime k^{\prime}_1}\cdots 
	U_d^{\prime k^{\prime}_d})=0, \tag{A.10}
\end{equation}

\noindent then, we can define the regularized integral as follows: 

\begin{equation}
	\int a\equiv \frac{1}{N}tr(a)=\frac{1}{N}a(0,0)\> tr1=a(0,0). 
	\tag{A.11}
\end{equation}

\noindent Using this integral, we can obtain the Yang-Mills theory on noncommutative space in the same way as the representation (A.1).

\vskip 1.5cm

%%%%%%%%%%%%%%%%%%%%%%%%%%%%%%%%%%%%%%%%%%%%%%%%%%%%%%%%%%%%%%%%%%%%%%%%%%%%%%%

\end{document}